\documentclass[]{interact}

\usepackage{epstopdf}
\usepackage[caption=false]{subfig}

\usepackage{multirow}

\usepackage[natbibapa,nodoi]{apacite}
\theoremstyle{plain}

\theoremstyle{definition}

\theoremstyle{remark}

\begin{document}


\title{Application of Multivariate Selective Bandwidth Kernel Density Estimation for Data Correction}

\author{
\name{Hai Bui \textsuperscript{a,b}\thanks{CONTACT: Hai Bui. Email: hai.bui@uib.no} and Mostafa Bakhoday-Paskyabi\textsuperscript{a,b}}
\affil{\textsuperscript{a}Geophysical Institute, University of Bergen, Allégaten 70, 5007 Bergen, Norway; \\ \textsuperscript{b}Bergen Offshore Wind Centre, Allégaten 55, 5007 Bergen, Norway}
}

\maketitle

\begin{abstract}
This paper presents an intuitive application of multivariate kernel density estimation (KDE) for data correction. The method utilizes the expected value of the conditional probability density function (PDF) and a credible interval to quantify correction uncertainty. A selective KDE factor is proposed to adjust both kernel size and shape, determined through least-squares cross-validation (LSCV) or mean conditional squared error (MCSE) criteria. The selective bandwidth method can be used in combination with the adaptive method to potentially improve accuracy. We demonstrate the efficacy of this approach using two examples: a hypothetical dataset and a realistic dataset involving wind measurements from the FINO1 meteorological mast station. The selective bandwidth methods consistently outperform non-selective methods, while the adaptive bandwidth methods improve results for the hypothetical dataset but not for the realistic dataset. The MCSE criterion minimizes root mean square error but may yield under-smoothed distributions, whereas the LSCV criterion strikes a balance between PDF fitness and low RMSE.
\end{abstract}

\begin{keywords}
Kernel Density Estimation; selective bandwidth; adaptive bandwidth; Bayesian data prediction
\end{keywords}

\section{Introduction}

The joint probability density function (PDF) describes the statistical relationship between two or more random variables. Among  its applications, the joint PDF can be used for an intuitive probabilistic prediction of a dependent variable (output) from one or several independent variables (inputs). Suppose we know the  joint PDF $f(X_1,...,X_N,Y)$ of $N$ inputs $X_1,...,X_N$ and an outputs $Y$, the conditional PDF can be computed as:

\begin{equation}
   f(Y \mid \tilde{X_1},...,\tilde{X_N}) =  \frac{f( \tilde{X}_1,...,\tilde{X}_N,Y)}{\int f( \tilde{X}_1,...,\tilde{X}_N,Y) dY },
   \label{eq_bayes}
\end{equation}

Here, ($\tilde{X}_1,...,\tilde{X}_N$) represents a specific realization of the input vector. In terms of Bayesian statistics, the conditional distribution discussed represents the posterior probability density function (PDF), which is an improvement upon the prior PDF distribution $f(Y)$. The conditional PDF provides valuable information about the output that can be inferred using various point estimation techniques, such as the conditional expectation $E(Y \mid X_1,...,X_N)$, or credible interval estimation \citep{edwards1963bayesian}. To illustrate this, consider a numerical weather model that predicts the temperature for the next 24 h to be 20$^\circ$C. However, we are aware that there are errors in the model's results. By constructing a joint PDF between the model predictions (inputs) and the observed temperature (output), we can derive a more accurate and useful prediction, for instance, the temperature is expected to be 22.5$^\circ$C $\pm$ 0.5$^\circ$C with 90\% certainty. Such a statistically informed correction can significantly enhance the reliability of forecasts and measurements alike.

The PDF is often unknown, but it can be estimated from observed data using either parametric or non-parametric methods. Parametric estimation relies on a theoretical statistical distribution and is applicable only in specific situations, making it unsuitable for complex distributions. Non-parametric estimation, on the other hand, employs methods like kernel density estimation (KDE), which do not assume a specific form for the PDF but instead use a kernel distribution function centered at each sample point \citep{izenman1991review}. One popular choice for the kernel function in non-parametric estimation is the Gaussian kernel function, which finds numerous applications \citep{chacon2018multivariate}. This method allows for more flexibility in estimating the joint PDF, making it useful in scenarios where the underlying distributions are non-Gaussian, multi-modal, or highly complex.

In the context of one-dimensional KDE, the variance of the kernel distribution is commonly referred to as the bandwidth. This bandwidth determines the size of the kernel and affects the smoothness of the density estimation \citep{chacon2018multivariate}. However, when considering the joint probability density function (PDF), the variance is replaced by the covariance matrix, making the estimated distribution dependent on both the size and shape of the kernel. In many applications, the covariance matrix of the kernel is computed by multiplying the sample covariance matrix by a scalar factor, also known as the bandwidth. This scalar factor determines the size of the kernel and significantly simplifies the computational complexity. The fixed bandwidth KDE method aims to find an optimal bandwidth, while the adaptive KDE method adjusts the bandwidth based on the density of the data \citep{abramson1982bandwidth,van2003adaptive}. However, both approaches maintain the same kernel shape. Although some argue that the kernel size is the most crucial factor, it is important to note that there may be situations where the kernel shape also plays an important role in the accuracy of the KDE method \citep{wand1993comparison,duong2003plug}.

This paper presents an intuitive method that addresses both the shape and size of the multivariate Gaussian KDE. We propose a selective KDE factor that enables flexible adjustments of the kernel size along each eigenvector of the covariance matrix. By selectively compressing or expanding the kernel, our method provides greater control over the estimation process. To facilitate practical implementation, we provide a Python implementation of the method as an extension of the Gaussian KDE class in the SciPy library \citep{2020SciPy-NMeth}. We demonstrate the effectiveness of our approach with two example applications. The first application involves hypothetical data, while the second utilizes real wind speed measurements obtained from a meteorological mast and LiDAR. Through these examples, we discuss the advantages and potential drawbacks of the proposed method.

\section{Methods}

\subsection{Multivariate Gaussian KDE}
Let's start by defining the inputs ($X_1, \ldots, X_N$) and the output $Y$ (as explained in Section 1) as a $(d=N+1)$-dimensional vector $\mathbf{X}=(X_1, \ldots, X_N, Y)$. Given a set of $M$ data points $[\mathbf{X}_1, \ldots, \mathbf{X}_M]$, the joint PDF $f(\mathbf{X})$ is a $d$-dimensional function that can be estimated using the KDE approach. The KDE estimate, denoted as $\hat{f}(\mathbf{X})$, is obtained by averaging a kernel function $K(\mathbf{X})$ that is evaluated at each data point:

\begin{equation}
\hat{f}(\mathbf{X})= \frac{1}{M} \sum_{i=1}^M K (\mathbf{X} - \mathbf{X}_i),
\label{eq_kde}
\end{equation}

In this paper, the chosen kernel function is the $d$-dimensional Gaussian kernel, given by:

\begin{equation}
K(\mathbf X; \mathbf H ) = \mathcal{N}(\mathbf X;0,\mathbf H) = \frac{1}{\sqrt{(2\pi)^d | \mathbf H |}} \exp \left(-\frac{1}{2} \mathbf X^\mathrm{T} \mathbf H^{-1} \mathbf X \right),
\label{eq_kernel}
\end{equation}

Here, the bandwidth matrix $\mathbf{H}$ is a positive definite, symmetric $d\times d$ matrix that determines the smoothing parameter for the KDE estimation.

\subsection{Fixed bandwidth (FW)}
 The selection of an appropriate bandwidth $\mathbf{H}$, which in this case is the covariance matrix of the Gaussian kernel, is crucial in the KDE method. One intuitive approach is to scale the covariance matrix $\mathbf{K}_{\mathbf{x}\mathbf{x}}$ of the sample (or training data) by a single bandwidth factor $h^2$:

\begin{equation}
\mathbf{H} = h^2 \mathbf{K}_{\mathbf{x}\mathbf{x}},
\label{eq_scalar_factor}
\end{equation}

Here, the scalar bandwidth $h$ depends on the sample size and the number of dimensions. When the joint PDF can be approximated by a unimodal distribution, plug-in choices such as Scott's Rule \citep{scott2015multivariate} with $h = M^{-1/(d+4)}$ or Silverman's Rule \citep{silverman1986density} with $h = [M(d+2)/4]^{-1/(d+4)}$ can be employed. However, for complex distributions, these plug-in rules often result in over-smoothed probability density functions (PDFs). In such cases, an optimal KDE factor $h$ can be determined using a bandwidth selection procedure (see Section \ref{sec_bw_selection}).

\subsection{Adaptive Bandwidth (AW)}
Using a fixed bandwidth can result in under-smoothing issues in regions with low data density. To address this, the kernel size can vary at each sample point, and Equation (\ref{eq_kde}) is modified as follows:

\begin{equation}
\hat{f}(\mathbf{X})= \frac{1}{M} \sum_{i=1}^M K_i (\mathbf{X} - \mathbf{X}_i),
\label{eq_kde_adaptive}
\end{equation}

Here, the kernel function $K_i(\mathbf{X})$ has a similar form to Equation (\ref{eq_kernel}), but the bandwidth matrix $\mathbf{H}$ is replaced with $\mathbf{H}_i$ defined as:

\begin{equation}
\mathbf{H}_i = \lambda_i^2 \mathbf{H},
\end{equation}

The local factor $\lambda_i$ depends on the sample density and is given by:

\begin{equation}
\lambda_i = \left[ \frac{\tilde{f}(\mathbf{X}_i)}{g} \right ]^{-\alpha},
\label{eq_localfactors}
\end{equation}

Here, $\tilde{f}(\mathbf{X}_i)$ is the first-guess density calculated at each data point using Equation (\ref{eq_kde}), $g$ is the geometric mean of $\tilde{f}(\mathbf{X}_i)$, and $\alpha \in [0,1]$ is a sensitivity parameter, which is typically set to 0.5 \citep{abramson1982bandwidth}. Consequently, $\lambda_i$ is larger in sparse regions and smaller in dense regions, allowing for adaptive smoothing of the KDE.

\subsection{Selective Bandwidth (SW)}
In the fixed and adaptive bandwidth methods, changing the parameter $h$ only alters the size of the kernel, while neglecting its shape. However, as demonstrated in subsequent sections, the shape of the kernel can be crucial in certain scenarios. To address this, we propose a selective bandwidth approach, utilizing a $d$-dimensional vector $\mathbf{h}=(h_1,..., h_d)$ to control the kernel's shape:

\begin{equation}
\mathbf{H} = \mathbf{Q [ (h^\mathrm{T} \mathbf{h}) \circ \Lambda ] Q^\mathrm{T}},
\label{eq_selective_factor}
\end{equation}

Here, $\circ$ denotes element-wise multiplication (Hadamard product), $\mathbf{Q}$ represents the matrix of eigenvectors, and $\Lambda$ is a diagonal matrix containing the eigenvalues obtained from the eigendecomposition of the sample covariance matrix ($\mathbf{K_{xx}} = \mathbf{Q \Lambda Q^\mathrm{T}}$).

Consequently, the selective factor $\mathbf{h}$ scales individual eigenvalues, enabling the kernel distribution to stretch or compress along each eigenvector's direction. When all $h_i=h$ for $i=1,...,d$, Equation (\ref{eq_selective_factor}) reduces to Equation (\ref{eq_scalar_factor}). Hence, the selective bandwidth method offers greater flexibility by allowing control over the kernel's shape. However, this enhancement increases the complexity of finding optimal parameters due to the higher degree of freedom. 

Table \ref{Table_KDE_methods} summarizes the available choices for KDE bandwidth along with their respective dependent parameters. Furthermore, the selective bandwidth method can be combined with the adaptive bandwidth method, resulting in the Selective-Adaptive Bandwidth method (SAW).

\begin{table}
\caption{\label{Table_KDE_methods}Choices of KDE bandwidth.}
\begin{center}
\begin{tabular}{ccc}
\hline
Method&Descripotion&Parameters\\
\hline
FW & Fixed bandwidth&  $h$\\
AW & Adaptive bandwidth &$h,\alpha$\\
SW & Selective bandwidth & $h_1,...,h_d$\\
SAW & Selective-Adaptive bandwidth &  $h_1,..,h_d,\alpha$\\
\hline 
\end{tabular}
\end{center}
\end{table}

\subsection{Bandwidth selection\label{sec_bw_selection}}

The optimal parameter(s) for each method can be determined by minimizing a criterion that evaluates the performance of the KDE. A commonly used criterion for this purpose is the least squared cross-validation (LSCV) \citep{gramacki2018nonparametric}, also known as unbiased cross-validation (UCV) \citep{chacon2018multivariate}. The LSCV criterion serves to assess the goodness-of-fit between the estimated distribution and the unknown true distribution. A smaller LSCV value indicates a closer match between the estimated distribution and the true distribution. The LSCV criterion is defined as follows:

\begin{equation}
\mathrm{LSCV}(\mathbf{H}) = \int_{\mathbb{R}^d} \hat{f}(\mathbf{X};\mathbf{H})^2 d\mathbf{X} - \frac{2}{M} \sum_{i=1}^M \hat{f}_{-i}(\mathbf{X}_i;\mathbf{H}),
\label{eq_LSCV}
\end{equation}

where $\hat{f}_{-i}(\mathbf{X})$ is the leave-one-out KDE with the data point $\mathbf{X}_i$ removed:

\begin{equation}
\hat{f}_{-i}(\mathbf{X})= \frac1{M-1} \sum_{j\neq i} K (\mathbf{X} - \mathbf{X}_j),
\label{eq_loo}
\end{equation}

The integration in Equation (\ref{eq_LSCV}) can be approximated using a simpler summation:

\begin{equation}
\int_{\mathbb{R}^d} \hat{f}(\mathbf{X};\mathbf{H})^2 d\mathbf{X} = \frac{1}{M}\sum_{i=1}^M \hat{f}(\mathbf{X}_i;\sqrt{2}\mathbf{H}).
\end{equation}

The computational complexity of Equation (\ref{eq_LSCV}) is $\mathcal{O}(n^2)$ for non-adaptive bandwidth methods. However, in the case of adaptive bandwidth methods, the second term on the right-hand side of Equation (\ref{eq_LSCV}) requires calculating the local bandwidth for each leave-one-out KDE, resulting in a computational complexity of $\mathcal{O}(n^3)$. This can be costly for large data samples. To reduce computational complexity, one can approximate LSCV by using the local bandwidth once for the entire data set.

In this study, the conditional expectation is employed as the prediction/correction for the output variable. Consequently, we test an additional criterion known as the mean conditional squared error (MCSE) to evaluate the performance. The MCSE measures the average squared error between the corrected $Y$ values and the corresponding values in the validation sample. Its definition is as follows:

\begin{equation}
\mathrm{MCSE}(\mathbf{H}) = \frac{1}{M} \sum_{i=1}^M \left[ \mathbf{E}\left(\hat{f}{-i}(Y|X{1,i},...,X_{N,i})\right) - Y_i \right]^2.
\label{eq_MCSE}
\end{equation}

In the case of a scalar KDE bandwith factor (FW and AW), where only a single parameter (e.g., $h$) needs to be determined, Equations (\ref{eq_LSCV}) and (\ref{eq_MCSE}) can be easily minimized numerically, for example using the golden section search method. Similarly, the Nelder-Mead method \citep{nelder1965simplex} (simplex downhill method) can be applied to our selective KDE bandwidth methods (SW and SAW).

\subsection{Python implementation}
We have developed a Python package called \texttt{sawkde} that implements the selective-adaptive Gaussian KDE method. The package has been published on Zenodo \citep{sawkde2023}. Additionally, we have made the code open-source and it is available on GitHub at the following URL: \url{https://github.com/haibuihoang/sawkde}. The package \texttt{sawkde} contains a class \texttt{saw\_gaussian\_kde} as an extension of the \texttt{gaussian\_kde} class from the SciPy package version 1.9.1 \citep{2020SciPy-NMeth}. The SciPy's \texttt{gaussian\_kde} class already has options to apply the bandwidth KDE factor $h$ using the Scott's or Silverman rule, and we added the following additional functions:

\begin{itemize}
  \item \texttt{set\_selective\_factor}: applies the selective factor (Equation \ref{eq_selective_factor}).
  \item \texttt{aw\_evaluate}: estimates the KDE using adaptive bandwidth (Equation \ref{eq_kde_adaptive}). This function needs to call the \texttt{calc\_local\_weights} function first to calculate the local factors (Equation \ref{eq_localfactors}).
\end{itemize}

A short tutorial on how to use \texttt{sawke} is included in the \texttt{sawkde} GitHub repository.

\section{Example 1: A hypothetical two-dimensional distribution}

To demonstrate the application of the KDE method for data prediction, we consider a simple relationship between an input variable $X$ and the true output variable $Y_t$ given by:

\begin{equation}
Y_t = \frac{X}{4} + \sin(X).
\end{equation}

\begin{figure}[ht]
\centerline{\includegraphics[width=0.8\linewidth]{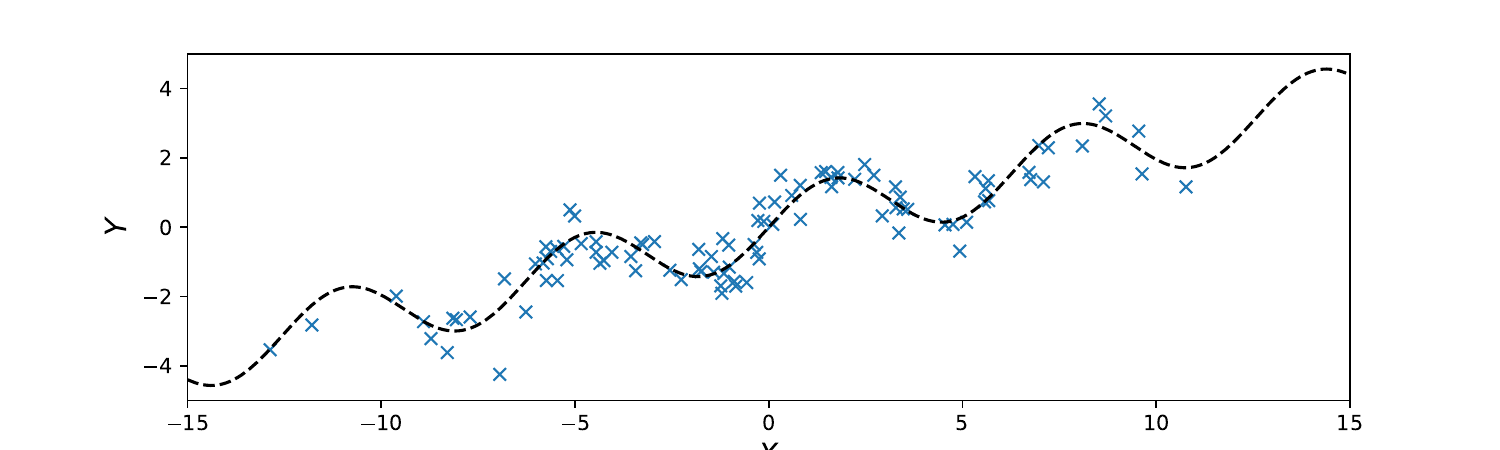}}
\caption{Sample data used in Example 1. The dashed line represents the target function $Y_t = X/4 + \sin(X)$. \label{fig_ex1_samples}}
\end{figure}

The target function consists of both a linear and a non-linear component, represented as the dashed line in Fig. \ref{fig_ex1_samples}. We assume that the input variable $X$ follows a Gaussian distribution $\mathcal{N}(0,5)$, and we measure the target function with an error term $\delta Y \sim \mathcal{N}(0,0.5)$. To estimate the target function, we generated a set of 100 2-dimensional data samples $(X, Y)$ as follows:

\begin{equation}
Y = \frac{X}{4} + \sin(X) + \delta Y,
\end{equation}

These generated samples are shown in Fig. \ref{fig_ex1_samples}. Similar to the regression approach, the expected values of the conditional expectation in Equation (\ref{eq_bayes}) should be capable of reconstructing a prediction that closely resembles the target function.

We constructed the 2-dimensional joint PDF$f(X,Y)$ Gaussian KDE with various methods: fixed bandwidth (FW), adaptive bandwidth (AW), selective bandwidth (SW), and selective-adaptive bandwidth (SAW). For the adaptive bandwidth methods (AW and SAW), we applied a smoothing parameter $\alpha=0.5$. Both the FW and AW methods employed a scalar bandwidth parameter $h$, which can be chosen using the Plugin method (Scott) or by minimizing the LSCV or MCSE criteria. On the other hand, the SW and SAW methods utilized a selective bandwidth parameter $\mathbf{h}=(h_1,h_2)$, also determined by minimizing the LSCV or MCSE criteria through a downhill search.

\begin{table}[htp]
\caption{\label{Table_KDE_methods_ex1}Bandwidth factors and associated criteria for different bandwidth methods in Example 1.}
\begin{center}
\begin{tabular}{cccc}
\toprule
Method & KDE Factor & LSCV($\times10^{-2}$) & MCSE($\times10$) \\
\midrule
\multicolumn{4}{c}{\emph{Plug-in (Scott) bandwidth}} \\
FW & 0.46 & -1.62 &  7.22 \\
AW & 0.46 & -1.72 &  7.21 \\
\midrule
\multicolumn{4}{c}{\emph{Optimized using LSCV}} \\
FW & 0.17 & -2.11 &  4.27 \\
AW & 0.17 & -2.27 &  4.21 \\
SW & [0.37 0.12] & -2.24 &  3.92 \\
SAW & [0.41 0.12] & \textbf{-2.37} &  3.85 \\
\midrule
\multicolumn{4}{c}{\emph{Optimized using MCSE}} \\
FW & 0.10 & -1.69 &  3.84 \\
AW & 0.10 & -1.68 &  3.84 \\
SW & [0.37 0.10] & -2.35 &  3.80 \\
SAW & [0.50 0.10] & -2.36 &  \textbf{3.78} \\
\bottomrule
\end{tabular}
\end{center}
\end{table}

\begin{figure}[ht]
\centerline{\includegraphics[width=0.8\linewidth]{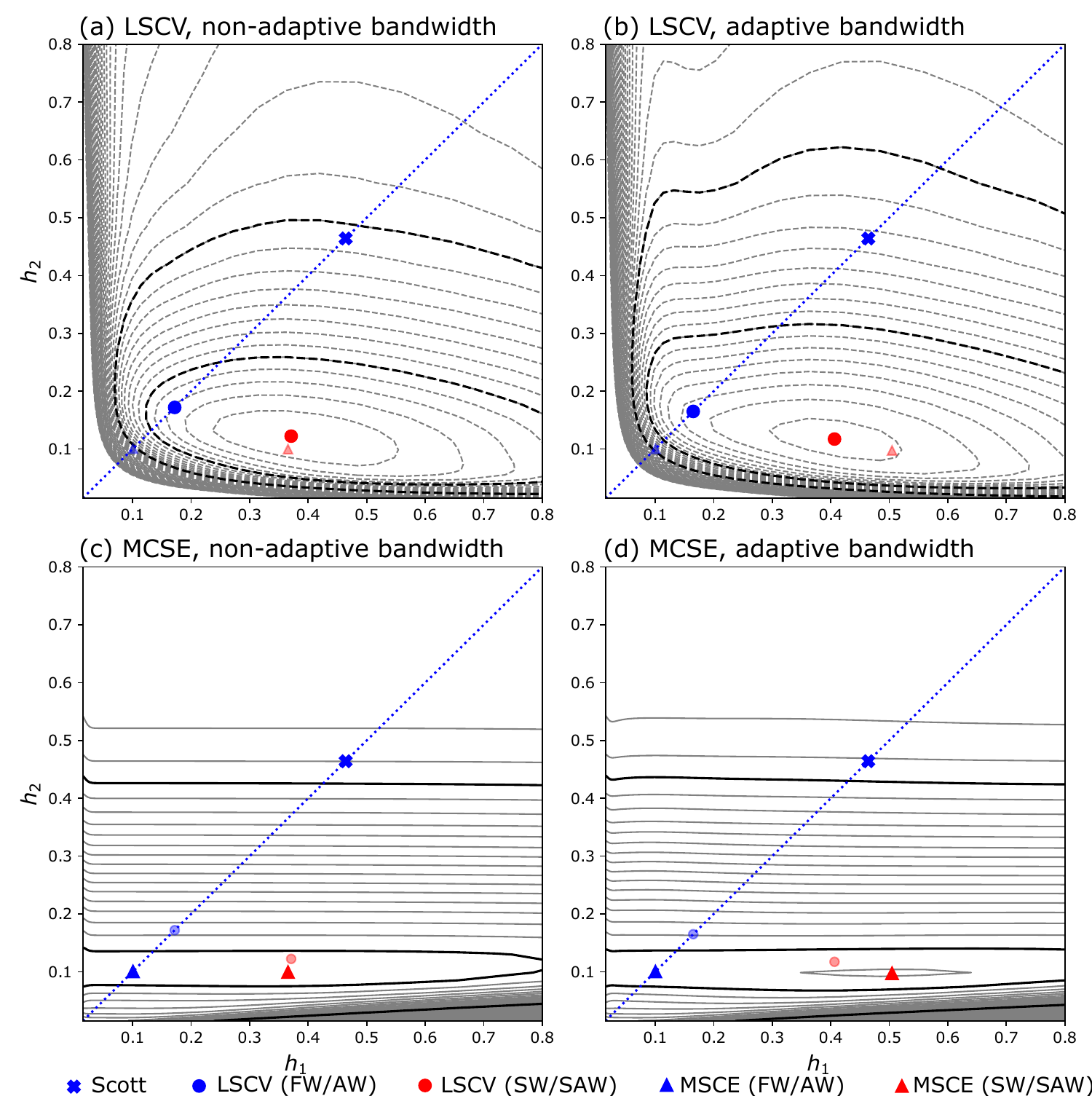}}
\caption{Contours of LSCV (a, b) and MCSE (c, d) in $(h_1,h_2)$-space for non-adaptive bandwidth methods (FW and SW) (a, c) and adaptive bandwidth methods (AW and SAW) (b, d). The blue markers on the diagonal lines ($h_1=h_2$) represent scalar bandwidth parameters (FW, AW), while the red markers indicate the optimal selective bandwidth parameters (SW, SAW). Circles (triangles) indicate optimal parameters determined using the LSCV (MCSE) criterion, and the blue cross represents the plug-in bandwidth obtained using the Scott method. The contour interval for LSCV (MCSE) is $5\times10^{-4}$ (0.02), and bold lines highlight contours at $-0.02$ (0.4) and $-0.016$ (0.8) for improved clarity.\label{fig_LSCV_MCSE}}

\end{figure}

Table \ref{Table_KDE_methods_ex1} provides the bandwidth parameters ($h$ or $h_1,h_2$) for different bandwidth methods. To visualize the relationships between these parameters, we have plotted them on the contour plot of LSCV and MCSE surfaces in Fig. \ref{fig_LSCV_MCSE}. In the figure, the scalar bandwidth factors (FW, AW) are represented by points on the diagonal lines ($h=h_1=h_2$). Upon initial observation, the structures of the LSCV and MCSE surfaces differ significantly, although they both exhibit similar global minima  ($h_1\approx0.45,h_2\approx0.1$). Because the global minima of these error surfaces generally do not occur on the diagonal lines, the performance of the selective bandwidth method is guaranteed to be better than or at least equal to that of the scalar bandwidth method, regardless of the criteria being used.  

\begin{figure}[ht]
\centerline{\includegraphics[width=0.8\linewidth]{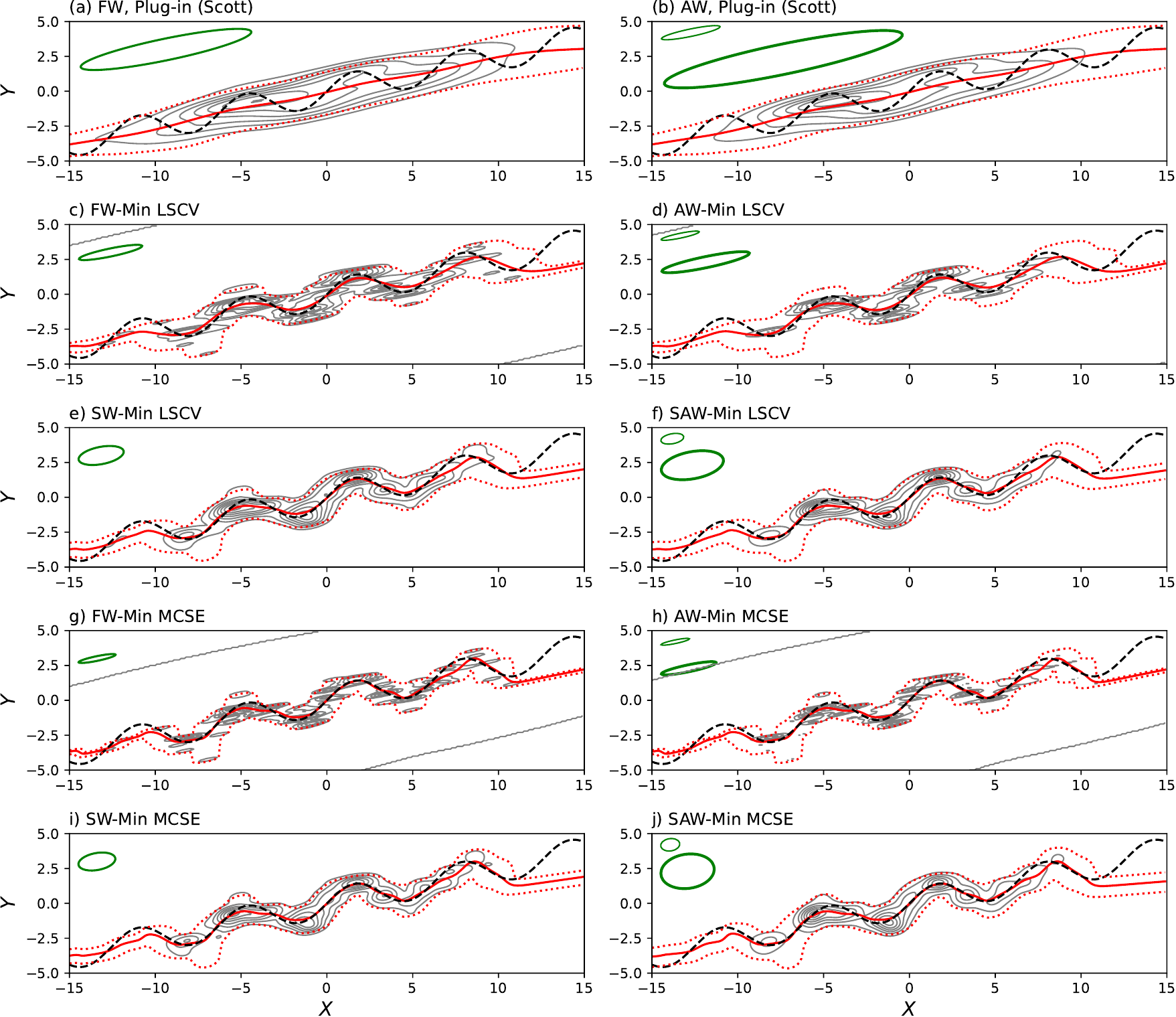}}
\caption{Comparison of Joint PDFs (gray contours), Conditional Expectations (red curves), and 90\% Credible Intervals (dotted red lines) using different bandwidth methods in Example 1. The black solid lines indicate the referenced target function. The green ellipses depict the shape and size of the kernels.\label{fig_ex1_jpdf}}
\end{figure}

The LSCV surfaces exhibit well-defined centers, indicating sensitivity to both $h_1$ and $h_2$ parameters (see Fig. \ref{fig_LSCV_MCSE}a, b). Conversely, the MCSE primarily depends on the $h_2$ parameter (see Fig. \ref{fig_LSCV_MCSE}c, d). Comparatively, the plugin bandwidth ($h=0.46$) is approximately three times larger than the optimal scalar bandwidth determined using the LSCV criterion ($h=0.17$), while the optimal scalar bandwidth obtained using the MCSE criterion is the smallest ($h=0.1$).
 
If the objective is to optimize the fitness of the distribution, the LSCV criterion is naturally preferred over the MCSE criterion. The plugin bandwidth ($h=0.46$) performs the worst for both adaptive (AW) and non-adaptive bandwidth (FW), resulting in the largest LSCV values (Table \ref{Table_KDE_methods_ex1}) and an over-smoothed joint PDF. However, when the LSCV criterion is employed, the bandwidth is reduced to approximately one third of the plugin value ($h=0.17$). Nevertheless, this reduction in bandwidth leads to under-smoothed features in the joint PDF (Fig. \ref{fig_ex1_jpdf}c, d). This is because the kernels assume a unimodal distribution based on the sample data, which has an elongated shape. However, by utilizing the selected bandwidth methods (SW, SAW), the kernel becomes less elongated, resulting in a better-fitted distribution (Fig. \ref{fig_ex1_jpdf}e, f) with smaller LSCV values. Among all the methods, the selective-adaptive method (SAW) performs the best.

If the objective is to optimize the fitness of the distribution, we may consider using the MCSE criterion. In such cases, the non-selective methods (FW, AW) result in even smaller bandwidth values ($h=0.1$), leading to heavily under-smoothed joint PDFs (Fig. \ref{fig_ex1_jpdf}g, h). Fortunately, the selective methods (SW, SAW) yield bandwidth factors that are quite close to those obtained using the LSCV criterion. However, due to the limited sensitivity of the MCSE criterion to $h_1$, there is a potential risk that minimizing based on this criterion could result in a value of $h_1$ that leads to either under-smoothed or over-smoothed joint PDFs in certain situations. Once again, the SAW method outperforms the others with the smallest MCSE value across all cases.

\section{Example 2: Wind Speed Correction}

\subsection{Data}
\begin{figure}[ht]
\centerline{\includegraphics[width=0.8\linewidth]{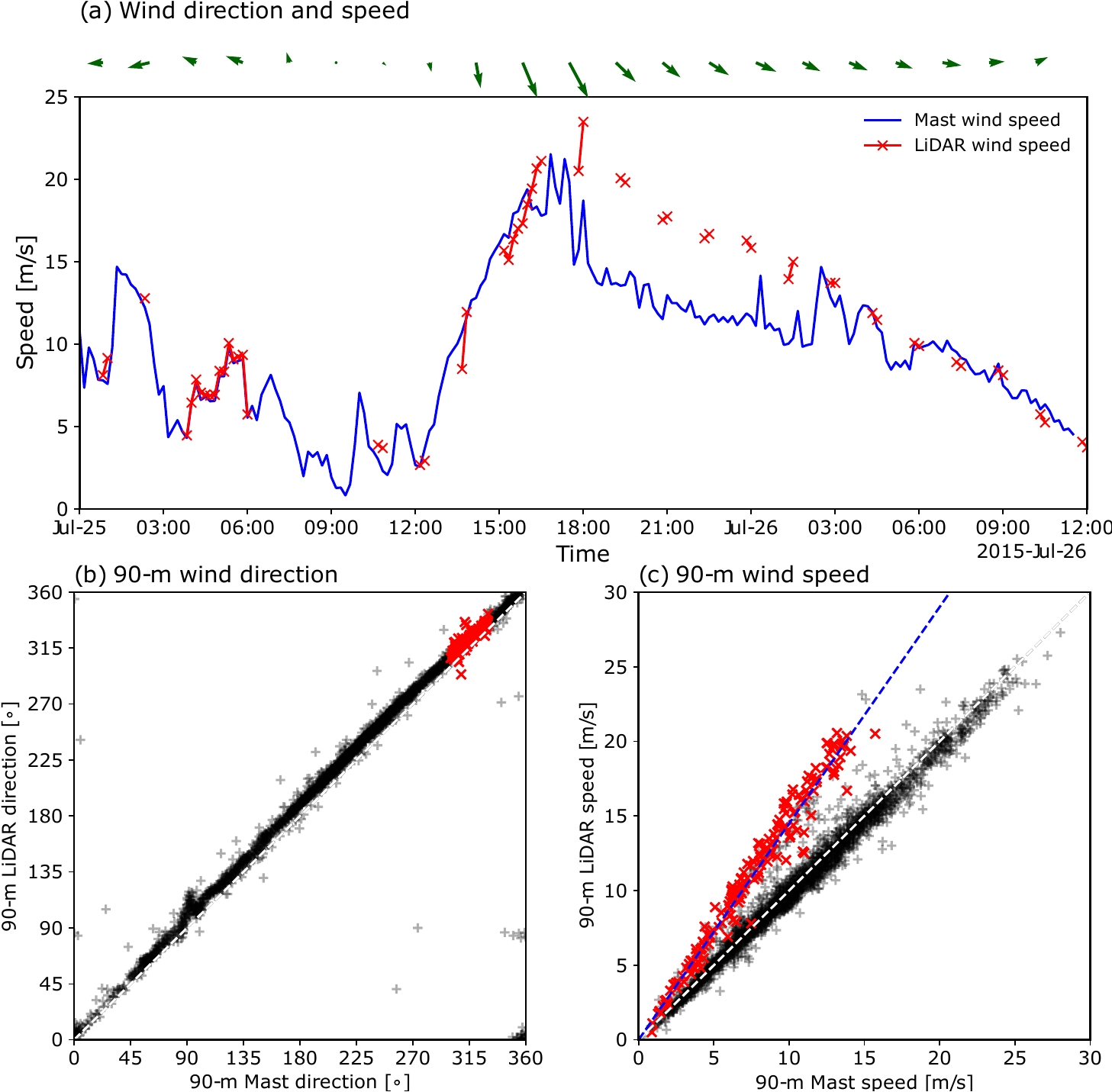}}
\caption{(a) Time series of wind speed from 00 UTC, July 25, 2015, to 12 UTC, July 26 of the 90-m cup anemometer (blue) at the FINO1 met-mast station and the LiDAR observation interpolated to 90 m (red), overlaid with green wind vectors. Scatter plots of wind directions (b) and wind speeds (c) of the FINO1 90-m cup anemometer and LiDAR data from May 22 to December 31, 2015. In (c), the data points with wind directions between 290$^\circ$ and 330$^\circ$ (i.e., around NW direction) are marked as red crosses, along with a regression line $y=1.45x$ (indicating that, on average, the LiDAR wind speed is 45\% higher than the cup wind speed or the mast wind speed is reduced by 31\%). \label{fig_ex2_data}}
\end{figure}

A meteorological mast is a tall tower with sensors attached to its side, designed to provide valuable measurements, particularly of wind speed within the planetary boundary layer. However, the presence of the mast tower causes flow distortion, leading to a shading effect on wind measurements obtained using anemometers. This shading effect is typically observed from a specific wind direction, known as the shadow zone \citep{orlando2011experimental,HIPERWIND_D21}. Rather than disregarding the measurements within this shadow zone, it is possible to correct the wind speed by considering the statistical characteristics of the shading effects.

In this section, we will demonstrate the wind speed correction using the joint PDF method with different KDE bandwidth methods proposed in the previous sections. To illustrate the correction process, we utilize wind measurements obtained from the cup anemometer at a height of 90 meters on the offshore FINO1 meteorological mast (Forschungsplattformen in Nord- und Ostsee No. 1). The mast is located at latitude   54.0149$^\circ$ N and longitude 6.5876$^\circ$ E in the German sector of the North Sea. Due to the shading effect caused by the mast itself, the cup anemometer measurements are affected when the wind blows from the northwest direction \cite{HIPERWIND_D21}.

To correct the mast wind speed, we employ wind measurements obtained using the Windcube\textsuperscript{\tiny\copyright} 100s wind LiDAR \citep{kumer2014comparison}. The LiDAR measures wind in a cone further away from the mast pole and is not subject to the shading effect \cite{HIPERWIND_D21}, making it suitable as a reference for correcting the mast anemometer measurements \citep{kim2019evaluation}. We utilized the LiDAR measurements interpolated to a height of 90 meters from May 22nd to December 31st, 2015. However, it should be noted that the LiDAR measurements are not continuous and contain numerous missing data points. Out of a total of 32,255 10-minute timesteps during the specified period, the interpolated LiDAR data at the 90-me level only comprises 3,421 data points (approximately 89.4\% missing), while the met mast measurements have 29,545 data points (8.4\% missing).

\subsection{Correction method}

Figure \ref{fig_ex2_data} illustrates a specific time period from 00Z on July 25th to 12Z on July 26th, in which the mast measurements are influenced by shading effects. The impact of shading is particularly pronounced between 18Z on July 25th and 06Z on July 26th. During this shading period, the wind speed measured by the mast is reduced by approximately 5 m/s compared to the wind speed measured by LiDAR. The shading zone is predominantly concentrated in the northwest direction, and on average, the wind speed reduction within this zone is approximately 31\% (Fig. \ref{fig_ex2_data}c). In contrast, the wind direction does not exhibit significant changes around the shading angle (Fig. \ref{fig_ex2_data}b).

Due to the significant influence of wind direction, relying solely on a 2-dimensional joint PDF of mast wind speed ($V_m$) and LiDAR wind speed ($V_L$) would not yield accurate corrections. Therefore, we aim to correct the wind speed using two inputs: the mast wind speed $V_m$ and the mast wind direction $\phi_m$. Specifically, we employ KDE methods to construct a 3-dimensional joint PDF $\hat{f}(X_1=V_m,X_2=\phi_m,Y=V_L)$. The expectation of the conditional PDF (Eq. \ref{eq_bayes}) is utilized as the corrected values, and we also calculate the 90\% credible interval based on the 5th and 95th percentiles of the conditional PDF.

After removing all data gaps, we obtained 3,057 data points from the LiDAR-mast measurements. These data points were divided into two sets: a training set comprising 2,446 data points (approximately 80\% of the data) and an independent validation set consisting of 612 points (20\% of the data). The validation set was randomly sampled from the original data, while the remaining data constituted the training set. The training set was used to establish the statistical relationships between the mast and LiDAR measurements (e.g. the joint PDFs), and the validation set was utilized to further evaluate the performance of each method using the root mean squared error (RMSE) as the evaluation metric.

\subsection{Results}

\begin{table}[htp]
\caption{\label{Table_KDE_methods_ex2} Bandwidth factors  ($h$ for scalar KDE factors and [$h_1,h_2,h_3$] for seletive KDE factors) and associated criteria for Example 2.}
\begin{center}
\begin{tabular}{ccccc}
\toprule
\multirow{2}{*}{Method} & \multirow{2}{*}{KDE Factor}  & \multicolumn{2}{c}{\emph{Train dataset}} & \emph{Test dataset} \\

\multicolumn{2}{c}{}  & LSCV$(\times10^{-4})$ & MCSE  &  RMSE\\

\midrule

Raw & & & &  1.27 \\
\midrule
\multicolumn{5}{c}{\emph{Plug-in (Scott) bandwidth}} \\
FW & 0.33  & -1.41 & 0.92 &   0.98 \\
AW & 0.33  & -1.39 & 1.02 &   1.03 \\
\midrule
\multicolumn{5}{c}{\emph{Optimized using LSCV}} \\
FW & 0.08  & -2.74 & 0.37 &   0.65 \\
AW & 0.08  & -2.58 & 0.38 &   0.65 \\
SW & [0.11 0.11 0.03]  & \textbf{-3.01} & 0.32 &  0.55 \\
SAW & [0.12 0.12 0.03]  & -2.88 & 0.32 &  0.56 \\
\midrule
\multicolumn{5}{c}{\emph{Optimized using MCSE}} \\
FW & 0.06  & -2.45 & 0.37 &   0.65 \\
AW & 0.06  & -2.26 & 0.37 &   0.65 \\
SW & [0.03 0.30 0.02]  & -2.25 & \textbf{0.30} &   \textbf{0.54} \\
SAW & [0.04 0.29 0.02]  & -2.26 & \textbf{0.30} &  \textbf{0.54} \\

\bottomrule
\end{tabular}
\end{center}
\end{table}

Similar to Table \ref{Table_KDE_methods_ex1}, Table \ref{Table_KDE_methods_ex2} provides an overview of the KDE bandwidth parameters for different methods along with their respective evaluation criteria. When comparing the corrected values to the raw data, the usage of the plugin bandwidth ($h=0.33$) reduced the RMSE from 1.27 m/s to approximately 1 m/s, resulting in an error reduction of around 21\%. In contrast to the idealized data in Example 1, where the adaptive methods demonstrated improved prediction performance, in this case, the fixed bandwidth (FW) approach performed slightly better with an RMSE of 0.98 m/s compared to the adapted bandwidth (AW) method with an RMSE of 1.03 m/s.

Upon applying the bandwidth selection techniques, the RMSE was further reduced to 0.65 m/s, corresponding to an error reduction of approximately 49\%, for both the FW and AW methods, irrespective of the LSCV or MCSE criteria. As expected, the selective bandwidth KDE methods (SW, SAW) achieved even lower RMSE values of around 0.55 m/s, resulting in an error reduction of approximately 57\%. However, once again, in terms of RMSE error, the adaptive (AW, SAW) methods did not outperform the non-adaptive bandwidth methods (FW, SW).

\begin{figure}[]
\centerline{\includegraphics[width=0.8\linewidth]{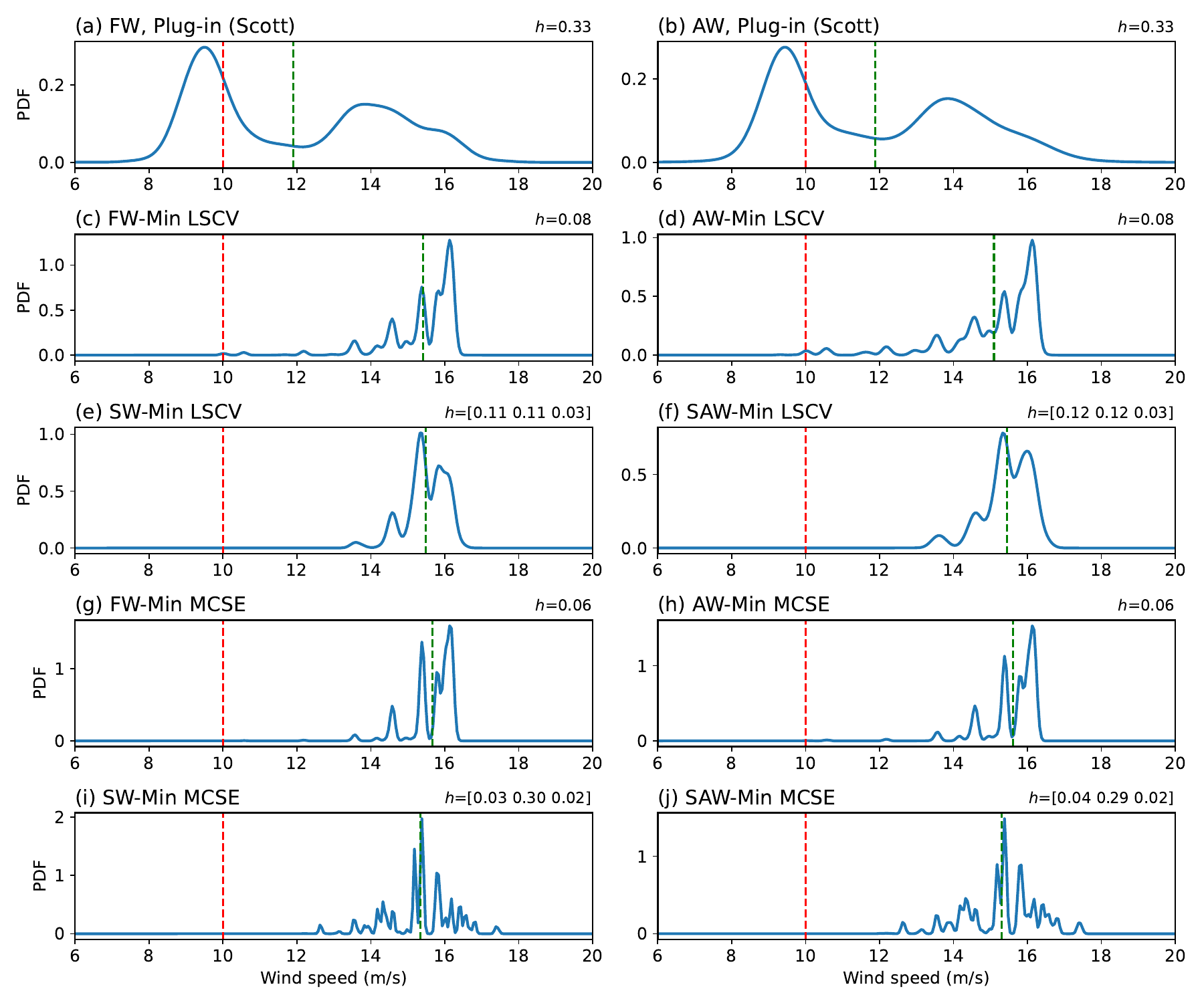}}
\caption{Conditional PDF $f(V_L | V_m=10,\phi_m=315^\circ)$ of Example 2 using different bandwidth methods. The dashed red lines indicate the speed of 10 m/s, and the dashed green lines represent the conditional expected values. \label{fig_ex2_cpdf}}
\end{figure}

Figure \ref{fig_ex2_cpdf} illustrates an example of the conditional PDFs $f(V_L|V_m=10,\phi_m=315)$ (i.e., in the northwest direction where the shading effect occurs) for different bandwidth methods. Assuming that the average value of the shading effect corresponds to a reference wind speed that is 45\% higher than the mast wind speed, the expected value of the conditional PDFs should be around 14.5 m/s (average correction). However, the plug-in bandwidth method produces smooth PDFs with two modal distributions: the primary mode around 9 m/s and the secondary mode around 14 m/s (Fig. \ref{fig_ex2_cpdf}a, b). The first mode arises from the data adjacent to the shading zone, while the second mode reflects the shading effect itself. In this case, the bandwidth along the wind direction dimension is too large, causing the data from the adjacent regions to dominate the information within the shading zone. As a result, the conditional expected value is too small (approximately 12 m/s) compared to the average correction.

When the LSCV criterion is applied for bandwidth selection, the resulting conditional PDFs become less smooth for all bandwidth methods (FW, AW, SW, SAW). As expected, the selective bandwidth methods (SW and SAW, Fig. \ref{fig_ex2_cpdf}e, f) exhibit smoother PDFs compared to the non-selective methods (FW and AW, Fig. \ref{fig_ex2_cpdf}c, d), indicating a reduction in under-smoothing issues. In addition to better fit (smaller LSCV), the selective bandwidth methods also demonstrate smaller MCSE and RMSE values compared to the non-selective methods (Table \ref{Table_KDE_methods_ex2}). The conditional PDFs generated using adaptive bandwidth (AW, SAW) appear slightly smoother than those using non-adaptive bandwidth (FW, SW).

When the MCSE criterion is applied, the focus shifts away from PDF fitness, resulting in clearly under-smoothed conditional PDFs, particularly noticeable for the selective bandwidth methods (SW and SAW, Fig. \ref{fig_ex2_cpdf}i, j). However, despite the under-smoothing, the conditional expected values obtained are closer to the average correction. Additionally, the selective bandwidth methods (SW and SAW) exhibit the best performance in terms of RMSE (Table \ref{Table_KDE_methods_ex2}).

\begin{figure}[htp]
\centerline{\includegraphics[width=0.9\linewidth]{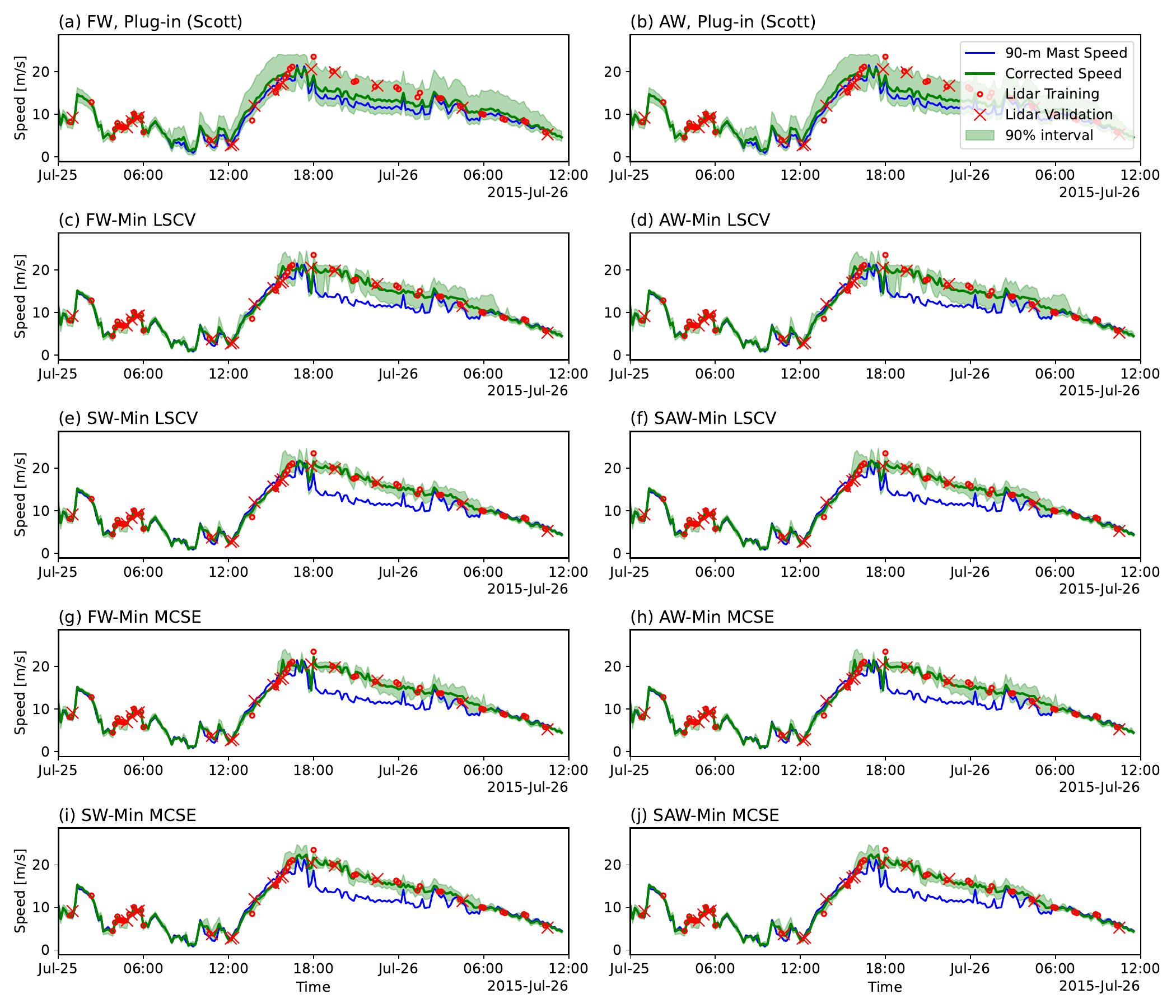}}
\caption{Time series of the mast wind speed (blue lines), corrected mast wind speed (green lines) with a 90\% credible interval (green shaded regions) for the example period from 00Z July 25 to 12Z July 2015. The LiDAR data used for training are shown as red circles, and the LiDAR data used for calculating RMSE are shown as red crosses. \label{fig_ex2_correctedspeed}}
\end{figure}

Figure \ref{fig_ex2_correctedspeed} shows an example of the corrected wind speed as well as the credible intervals, which represent the uncertainty of the correction, using different methods for the same period as Fig. \ref{fig_ex2_data}a. The plugin methods (\ref{fig_ex2_correctedspeed}a, b) perform the worst with the wind speed during the shadow period being only corrected slightly. The other optimal bandwidth methods perform well to bring up the erroneous mast wind speed during the shading period to the same level as the mast LiDAR speed with a much narrower region of credible interval. Although being visually similar, the corrected speed using the selective bandwidth methods (\ref{fig_ex2_correctedspeed}e,f,i,j) performs better with a smaller uncertainty interval compared to the non-selective bandwidth methods (\ref{fig_ex2_correctedspeed}c,d,g,h). The difference between the adaptive and non-adaptive methods is visually insignificant. 

Another advantage of utilizing conditional PDFs is the robustness of the credible interval, which provides valuable information about the reliability of the corrected values. The width of this interval serves as an indicator of the uncertainty associated with the corrections. In the case of the plug-in bandwidth method, a wide credible interval suggests poorer performance, indicating higher uncertainty in the corrected values. Furthermore, both before and after the shadow period, the credible interval is notably narrower, indicating measurement error being small for the wind speed of the mast anemometer.

\section{Conclusions and remarks}

This paper presents an intuitive application of multivariate kernel density estimation (KDE) for data correction. The data is corrected using the expected value of the conditional probability density function (PDF) along with a credible interval that quantifies the uncertainty in the correction. The conditional PDF is computed based on the joint PDF estimated using multivariate Gaussian KDE. In addition to using a scalar KDE factor, we proposed a selective KDE factor that can not only adjust the size but also the shape of the kernel. The optimal selective KDE factor can be determined by minimizing either the least-squares cross-validation (LSCV) or the mean conditional squared error (MCSE) criteria. Additionally, for potential performance improvement, the selective bandwidth method can be combined with the adaptive bandwidth method, where the bandwidth is larger in data regions with lower density.

We provide two examples of the method's application. The first example utilizes a hypothetical dataset, while the second example involves real measurements of 90-m wind at the FINO1's meteorological mast station, which is significantly influenced by the mast pole and experiences shading effects at specific wind directions. In both examples, the default (plug-in) bandwidth does not perform well for such complex distributions, resulting in over-smoothed distributions due to the excessively large bandwidth. By employing the bandwidth selection method using LSCV, the non-selective bandwidth method outperforms the plug-in bandwidth method but may yield under-smoothed PDFs. On the other hand, the selective bandwidth method consistently demonstrates the best performance, whether the objective is the PDF fitness (using LSCV) or the error metrics (MCSE or RMSE). The adaptive bandwidth method, however, does not consistently improve the accuracy of the correction. While it performs slightly better than the non-adaptive method for the hypothetical dataset, the effect is either insignificant or slightly worse for our realistic dataset.

Lastly, between the two criteria for bandwidth selection, the MCSE criterion minimizes the root mean square error but can result in an under-smoothed joint distribution. If the smoothness of the PDF needs to be considered, the LSCV criterion can provide a good compromise between a better-fitted PDF and a small RMSE of the corrected data.


\section*{Data Availability}
Our implementation of the selective-adaptive Gaussian KDE method in Python/Cython has been recently published, along with a tutorial that can be downloaded from the following link: \url{https://doi.org/10.5281/zenodo.7896366} \citep{sawkde2023}. Additionally, we have made the code available as an open-source project on GitHub: \url{https://github.com/haibuihoang/sawkde}.

\section*{Competing interests} 
All authors declare that they have no competing of interest.

\section*{Acknowledgments}
The work was supported by the CONWIND project (Research on smart operation control technologies for offshore wind farms), ENERGIX program, Grant number 304229. LiDAR data used in this study were gathered as part of the OBLEX-F1 field campaign that has been performed under the Norwegian Centre for Offshore Wind Energy (NORCOWE). The FINO1 meteorological reference data were provided by Deutsches Windenergi Institute (DEWI). 



\bibliographystyle{apacite}

\bibliography{main}

\end{document}